# Thermodynamic aspects of materials' hardness: prediction of novel superhard high-pressure phases

V.A. MUKHANOV, O.O. KURAKEVYCH and V.L. SOLOZHENKO[*]

*LPMTM-CNRS, Université Paris Nord, Villetaneuse, France*

In the present work we have proposed the method that allows one to easily estimate hardness and bulk modulus of known or hypothetical solid phases from the data on Gibbs energy of atomization of the elements and corresponding covalent radii. It has been shown that hardness and bulk moduli of compounds strongly correlate with their thermodynamic and structural properties. The proposed method may be used for a large number of compounds with various types of chemical bonding and structures; moreover, the temperature dependence of hardness may be calculated, that has been performed for diamond and cubic boron nitride. The correctness of this approach has been shown for the recently synthesized superhard diamond-like $BC_5$. It has been predicted that the hypothetical forms of $B_2O_3$, diamond-like boron, $BC_x$ and $CO_x$, which could be synthesized at high pressures and temperatures, should have extreme hardness.

**Keywords:** superhard materials, theory of hardness, thermodynamical properties

* e-mail: vladimir.solozhenko@univ-paris13.fr

---

[*] Corresponding author. Email : vls@lpmtm.univ-paris13.fr

The synthesis of man-made superhard phases started from the early 50's, as soon as development of high-pressure techniques allowed reaching the pressures necessary for diamond synthesis. Nevertheless, the theoretical design of novel superhard materials is a great challenge to materials scientists till now. Hardness describes the abrasive properties of materials and is understood as the ability of a material to resist an elastic and plastic deformation or brittle failure [1-3]. Till present time many attempts to predict materials' hardness have been made using the structural data and such materials' characteristics as bulk ($B$) and shear ($G$) moduli, specific bond energy, band gap ($E_g$), density of valent electrons (i.e. the number of valent electrons per unit volume $N_e$), etc. [1-9], however, all the proposed models either are very complicated and suggest performing the *ab initio* calculations, or do not allow the satisfactory description of crystals with various bonding type and prediction of hardness of hypothetical phases.

The universal model of hardness should also take into account the microstructure of materials (grain size, inter-grain boundaries, etc.) [3,10,11]. However, these factors are usually ignored in theoretical simulations, so that the calculated values correspond to so-called "chemical" hardness that is usually observed only for single crystals and well-sintered polycrystalline bulks. Here we will deal with the "chemical" hardness only.

Up to date the best correspondence between the calculated and experimental values of hardness has been achieved in the recent papers [5,6]. According to [5],

$$H_V = 350 \frac{N_e^{2/3} e^{1.191 f}}{d^{5/2}} \tag{1}$$

where $H_V$ – hardness in GPa, $N_e$ – electron density, $f$ – ionicity of bonding, $d$ – bond length. In paper [6] another formula for hardness calculation has been proposed, i.e.

$$H_V = \frac{C}{\Omega} \frac{\sqrt{e_X e_Y}}{d_{XY} n_{XY}} e^{-\sigma f} \tag{2}$$

where $C = 1550$ and $\sigma = 4$ are empirical coefficients, $\Omega$ – volume of atoms' couple $XY$, $n$ – number of bonds between atom $X$ and neighboring atoms $Y$, $e_j = Z_j / R_j$, where $Z_j$ – number of valent electrons of atom $j$ ($X$ or $Y$), and radius $R_j$ is chosen in such a way that the bounded sphere contains exactly $Z_j$ valent electrons. However, in both cases [5,6] the calculations are rather difficult, while the final accuracy is about 10%. As soon as the problem of the theoretical estimation of hardness is not unambiguously resolved even for the known compounds, the results of calculations for hypothetical compounds should be taken with great precaution.

In order to predict novel superhard phases, the *ab initio* calculations of bulk modulus $B = dp/d\ln V$ are widely used [7,12], because the hard phases usually are low compressible [13-15]. However, the employed methods for the calculation of $B$ are complicated and ambiguous, while the bulk modulus itself is not a good predictor of hardness [16].

The purpose of present work was to establish a simple quantitative dependence of hardness and compressibility of solids on there structural and thermodynamic properties.

According to our concept, the hardness of a phase is proportional to the atomization energy, which may be considered as a characteristic of the bond rigidity (for the clarity, we will use the standard values of Gibbs energy of atomization $\Delta G°_{at}$), and is in inverse proportion to the molar volume of a phase[1] and to the maximal coordination number of the atoms. The value defined in such way has the dimensions of pressure. The plasticity of materials is taken into account by the empirical coefficient $\alpha$. In general case the polarity of bonds leads to the hardness decrease, which may be clearly seen in the sequence of isoelectronic analogues of diamond, i.e. diamond (115 GPa [18]) – cubic boron nitride cBN (62 GPa [19]) – BeO (13 ГПа [3-5]) – LiF (1.5 ГПа [3-5]). This factor has been evaluated by empirical coefficient $\beta$, which is the measure of the bond covalency.

The formula that allows calculating the Vickers hardness ($H_V$) of crystals at 298 K is

$$H_V = \frac{2\Delta G°_{at}}{VN}\alpha\beta\varepsilon, \qquad (3)$$

where $V$ – molar (atomic) volume (cm$^3$ mole$^{-1}$); $N$ – maximal coordination number; $\alpha$ – coefficient of relative (as compared to diamond) plasticity; $\beta$ – coefficient corresponding to the bond polarity (see below); $\varepsilon$ – ratio between the mean number of valent electrons per atom and the number of bonds with neighboring atoms ($N$)[2]; $\Delta G°_{at}$ – standard Gibbs energy of atomization (kJ mole$^{-1}$) of compound $X_mY_n$.

$$\Delta G°_{at\ X_mY_n} = m\Delta G°_{at\ X} + n\Delta G°_{at\ Y} - \Delta G°_{f\ X_mY_n}; \qquad (4)$$

where $\Delta G°_{f\ X_mY_n}$ – standard Gibbs energy of formation of $X_mY_n$, $\Delta G°_{at\ X}$ and $\Delta G°_{at\ Y}$ – standard Gibbs energy of atomization of elements $X$ и $Y$.

Coefficient $\alpha$ has been estimated from the experimental values of $H_V$ for diamond, d-Si, d-Ge and d-Sn. For the elementary substances and compounds of second period elements $\alpha$ equals 1, while for other periods ($\geq$3) it makes 0.7. This coefficient reflects the difference in the bond strength [6] for the elements of different periods.

Coefficient $\beta$ (square of the covalency $f$) has been calculated by the equation

$$\beta = \left(\frac{2\chi_Y}{\chi_Y + \chi_X}\right)^2; \qquad (5)$$

where $\chi_X$, $\chi_Y$ – electronegativities of the elements by Pauling, $\chi_X > \chi_Y$ [21]. For elementary substances $\beta = 1$.

For the refractory crystalline compounds the values of hardness calculated by equation (3) are in a very good agreement (less than 4 GPa of discrepancy, i.e. < 7%) with the experimental values [4-6,18-35] (Fig. 1a).

The proposed method also allows to calculate the values of hardness at various temperatures by introducing the linear approximation of temperature dependence of $\Delta G_{at}(T)$, i.e.

---

[1] E.g. for carbon phases the linear dependence between the hardness and density has been established in paper [17].
[2] The use of this coefficient allows to establish the hardness of the compounds A$^I$B$^{VII}$ ($\varepsilon = 1/N$) and A$^{II}$B$^{VI}$ ($\varepsilon = 2/N$), i.e. LiF, NaCl, BeO, ZnS, MgO, etc.

$$\Delta G_{at}(T) = \Delta G_{at}(300) \cdot [1-(T-300)/(T_{at}-300)], \qquad (6)$$

where $T_{at}$ – temperature of atomization[3]; as well as by introducing the temperature dependences of molar volumes $V(T)$. The resulting equation is

$$H(T) = H(300) \cdot \frac{\Delta G_{at}(T) \cdot V(300)}{\Delta G_{at}(300) \cdot V(T)}. \qquad (7)$$

Fig. 2 shows the temperature dependences of Vickers and Knoop hardness for diamond and cBN in comparison with experimental data [22,23]. The theoretical curves reasonably describe the experimentally observed hardness decrease with temperature. At relatively high temperatures (~ 0.3-0.5 $T_{at}$) this equation gives 10-15% higher values than observed ones, that should be attributed to the increase of materials' plasticity due to the intensification of the surface and bulk diffusion [36].

One more advantage of the proposed method is the possibility to estimate the hardness of various forms of boron and its compounds ($B_4C$, $B_6O$, $B_{13}N_2$), that is rather complicated by using other methods because of extreme complexity of boron-related structures. In our calculations we have taken the mean value of electronegativities of all atoms connected to $B_{12}$ icosahedron as a $\chi$ value for anion. Thus, the calculated values of Vickers hardness for $B_4C$ and $B_6O$ are 44 and 38 GPa, respectively; that is in a very good agreement with the experimental data for single crystal $B_4C$ ($H_V$ = 45 GPa [24]) and polycrystalline $B_6O$ ($H_V$ = 38 GPa [25]). The lower value of than that of B-C bonds. The estimation of hardness for the recently synthesized rhombohedral boron subnitride $B_{13}N_2$ [37] has given $H_V$ = 40.3 GPa[4] that allows ascribing $B_{13}N_2$ to superhard phases.

In the framework of the proposed model it is possible to calculate the hardness of phases that have not been synthesized to present time, e.g. $C_3N_4$ with $Si_3N_4$ structure [7], $CO_2$ with α-$SiO_2$ structure, hp-$B_2O_3$ with $Al_2O_3$ structure [38] and diamond-like phases of the B–C system [12,39] (see Table 1). In all cases the molar volumes have been calculated from the covalent radii of the elements, while $\Delta G°_f$ values of the phases have been fixed to the standard Gibbs energies of formation of known compounds in the corresponding binary systems ($C_2N_2$, $CO_2$, $B_4C$, β-$B_2O_3$ [22,26-30,40]). The applicability of this method for estimating the hardness of hypothetical compounds has been recently illustrated by the example of diamond-like $BC_5$ (c-$BC_5$), a novel superhard phase synthesized under high pressures and temperatures [41]. Vickers hardness of this phase has been estimated as 70.6 GPa (Table 1), which is in excellent agreement with the experimental value $H_V$ = 71(8) GPa [41].

In the framework of our approach, the compressibility $K$ of a phase at 298 K is proportional to the molar volume $V$ and is in inverse proportion to Gibbs energy of atomization $\Delta G°_{at}$ (see, for example [42]), so

$$K = g \frac{V}{3f\Delta G°_{at}}, \qquad (8)$$

---

[3] For diamond and cBN the corresponding temperatures of sublimation are 4300 K and 3300 K, respectively [22].

where $f = \sqrt{\beta} = \dfrac{2\chi_Y}{\chi_X + \chi_Y}$ – covalency of chemical bonds. The empirical coefficient "3" in equation (8) has been established from the experimental data on the compressibility of cBN, d-Si and d-Ge [43,44], while $g$ is some correction coefficient (see below). For the majority of the closely packed covalent compounds and metals there is a good agreement between the values of $K_{exp}$ и $K_{theor}$, however, for the phases with anisotropic lattices, alkali and some alkali-earth metals the calculated values are lower than the experimental ones. For transition metals of periods V and VI, $g = 0.625$ should be introduced in formula (8). Fig. 1b shows the comparison between experimental and theoretical values of bulk modulus for various compounds. The remarkable deviation ($K_{exp}/K_{theor} \sim 1.5$) is observed only for three most hard phases, i.e. diamond, cubic $BC_2N$ and diamond-like $BC_5$.

By combining equations (3) and (8), obtain

$$H_V = \dfrac{2}{3}\dfrac{g\alpha\varepsilon\sqrt{\beta}}{N}B \qquad (9)$$

that illustrates the famous correlation between bulk modulus $B$ (the value inverse to compressibility $K$) and hardness $H_V$ [13-15].

Thus, in the present work it has been shown that hardness and compressibility of solids are directly related to the thermodynamic and structural properties. The formulated equations may be used for a large number of compounds with various types of chemical bonding and structures. In the framework of proposed method we have calculated the temperature dependencies of hardness for diamond and cubic boron nitride. Our method allows estimating the hardness and compressibility of various hypothetical compounds using the data on Gibbs energy of the atomization of elements and covalent/ionic radii. The capacity of this approach to predict hardness has been illustrated by an example of the recently synthesized superhard diamond-like $BC_5$ [41].

The authors are grateful to the Agence Nationale de la Recherche for financial support (grant ANR-05-BLAN-0141).

---

[4] The $2\Delta G°_{at}/NV$ value has been set to a mean (~51 GPa) of corresponding values for $B_6O$ and $B_4C$; $\beta = 0.79$.

Table 1.

Theoretical values of microhardness for some hypothetical superhard high-pressure phases

| Solids* | $-\Delta G°_f$, kJ mole$^{-1}$ [22,26-30] | $-\Delta G°_{at}$, kJ mole$^{-1}$ [21,22, 26-30] | $V$, cm$^3$ mole$^{-1}$ | $N$ | $\dfrac{2\Delta G°_{at}}{NV}$, GPa | $\chi_X$ [21] | $\chi_Y$ [21] | $\beta$ | $H_{V\,\text{theor}}$, GPa, (3) |
|---|---|---|---|---|---|---|---|---|---|
| C$_3$N$_4$ | 60† | 3896.0 | 35.45† | 4 | 55.0 | 3.04 | 2.55 | 0.8393 | 41.7 |
| c-BC$_5$ | 96† | 3971.1 | 21.32 | 4 | 93.1 | 2.55 | 2.04 | 0.7903 | 70.6 |
| c-BC$_3$ | 62† | 2594.6 | 14.09† | 4 | 92.1 | 2.55 | 2.04 | 0.7903 | 73.2 |
| d-B‡ | 0† | 518.8 | 4.242‡ | 4 | 61.2 | 2.04 | 2.04 | 1 | 61.2 |
| hp-B$_2$O$_3$ | 1272.9† | 3005.7 | 22.29§ | 4 | 67.4 | 3.44 | 2.04 | 0.5543 | 37.4 |
| hp-B$_2$O$_3$ | 1272.9† | 3005.7 | 21.0** | 6 | 47.07 | 3.44 | 2.04 | 0.5543 | 26.4 |
| CO$_2$ (α-SiO$_2$) | 294.0† | 1429.0 | 14.5† | 4 | 49.3 | 3.44 | 2.55 | 0.725 | 35.7 |
| "d-C$_2$O"†† | 148.7† | 1722.9 | 10.64 | 4 | 81.0 | 3.44 | 2.55 | 0.725 | 58.7 |
| d-CO | 37.0† | 940.0 | 5.90 | 4 | 79.7 | 3.44 | 2.55 | 0.725 | 57.8 |

\*    the calculations have been performed with ε = 1 and α = 1.

†    the values have been estimated using the standard Gibbs energies of formation of known compounds in the corresponding binary systems;

‡    the length of B–B bond taken as 1.66 Å;

§    molar volume of β-B$_2$O$_3$ phase;

**    estimation for the lowest possible limit of the molar volume of B$_2$O$_3$ according to the covalent radius data [21];

††    buckled layers of graphite are connected by oxygen atoms.

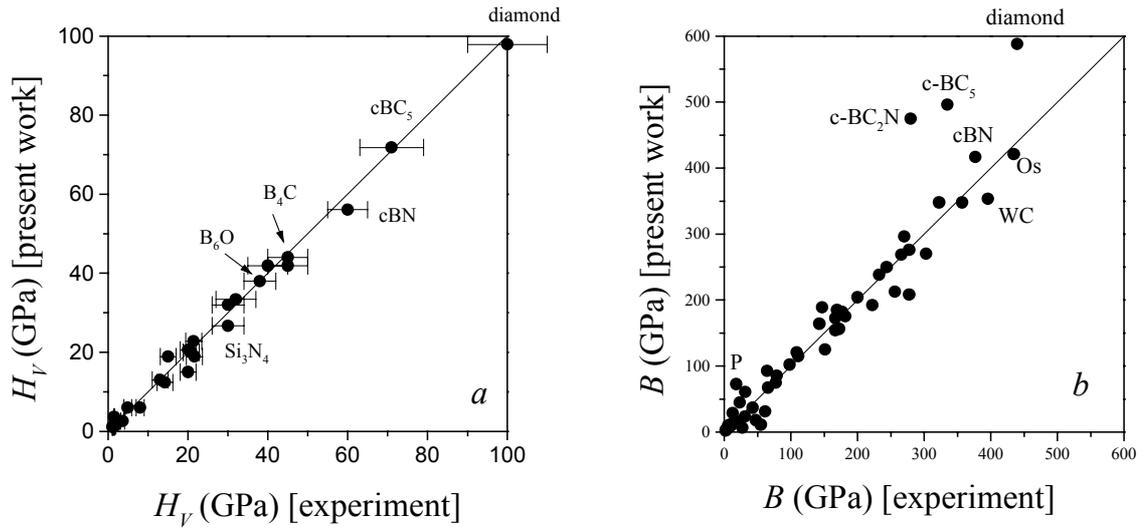

Fig. 1 (*a*) Comparison of experimental values of Vickers hardness of various phases with corresponding values calculated in the framework of model proposed in the present paper. (*b*) Comparison of experimental bulk moduli of various phases with values calculated by equation (8).

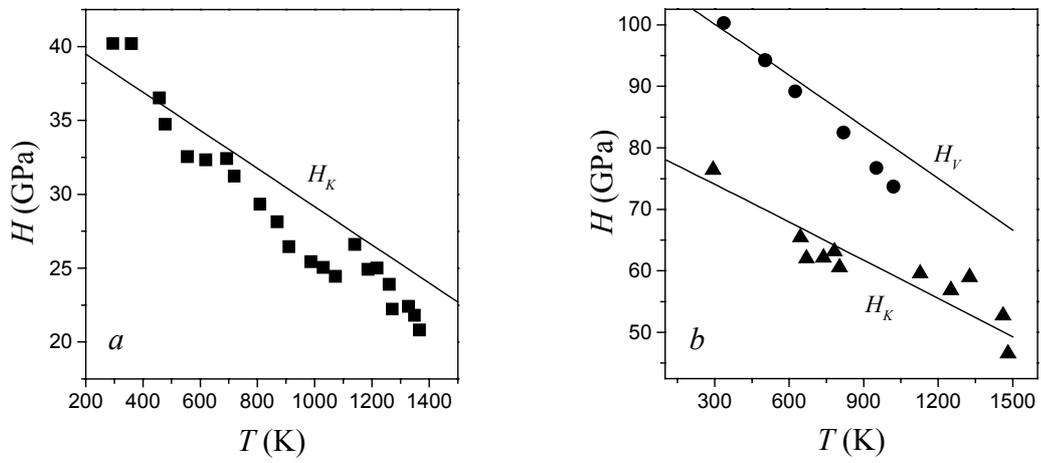

Fig. 2 Temperature dependence of hardness of polycrystalline (mean particle size of 5 mkm) cBN (*a*) and single-crystal diamond (*b*). The lines correspond to the results of calculation by equation (7). The symbols represent the experimental data obtained by static indentation [22,23].